\documentstyle[twoside,fleqn,espcrc2,epsfig]{article}

\newcommand{\la}{\raise.16ex\hbox{$\langle$}}
\newcommand{\ra}{\raise.16ex\hbox{$\rangle$}}
\newcommand{\be}{\begin{equation}}
\newcommand{\ee}{\end{equation}}
\newcommand{\bea}{\begin{eqnarray}}
\newcommand{\eea}{\end{eqnarray}}

\title{The Static Baryon Potential}

\author{
C.~Alexandrou\address{Department of Physics, University of Cyprus, 
CY-1678 Nicosia, Cyprus}%
\thanks{Talk presented by C.~Alexandrou.},
Ph.~de~Forcrand\address{Institut f\"ur Theoretische Physik, ETH-H\"onggerberg, CH-8093 Z\"urich,  Switzerland and Theory Division, CERN, CH-1211 Geneva 23, Switzerland},
A.~Tsapalis\address{Department of Physics, University of Athens, Athens, Greece}\thanks{Acknowledges funding from the 
University of Cyprus and the European network ESOP (HPRN-CT-2000-00130).}
}

\begin{document}

\begin{abstract}
Using state of the art lattice techniques we investigate the static
baryon potential. We employ the multi-hit procedure for the time links and
a variational approach to determine the ground state with sufficient
accuracy that, for distances up to $\sim 1.2$~fm,
 we can distinguish the $Y$- and $\Delta$- Ans\"atze for the baryonic 
Wilson area law. Our analysis shows that the $\Delta$-Ansatz is favoured.
This result is also 
supported by the gauge-invariant nucleon wave function which we measure
for the first time.
\end{abstract}

\maketitle

\vspace*{-10.5cm}
\begin{flushleft}
CERN-TH/2001-317
\end{flushleft}

\vspace*{8.cm}

\section{Introduction}
Due to  the 
 important role which the  $q\bar{q}$ potential plays in our understanding
of  the structure of mesons, there exist numerous studies of this quantity
on the lattice. For a review see ref.~\cite{Bali}.
The three quark potential plays an equally important role in the
understanding of baryon structure. Mass relations
between baryons and mesons can be made more exact if the form of the
baryonic potential is known~\cite{meson-baryon}.
However, very few lattice studies have been made of the baryonic potential. 
Moreover, two such recent studies~\cite{Bali,Takahashi} have
reached different conclusions for the area law behaviour of the
baryonic Wilson loop: they give support to two different Ans\"atze,
called $Y$- and $\Delta$-law.
Since 
the maximal
difference between these
is a mere  15\% for $SU(3)$, a reliable extraction of the ground state 
as well as noise reduction techniques are essential in order 
to resolve the dominant area law behaviour~\cite{AFT}.

In this work, employing state of the
art lattice techniques,  we are  able to reach sufficient
accuracy to distinguish between the two
Ans\"atze up to distances of $\sim 1.2$~fm. At the same time,
we compare our lattice results directly to the sum of $q\bar{q}$ potentials
measured on the same lattices, thus avoiding any model assumptions.
We also compare our procedure and  results 
with those of ref.~\cite{Takahashi} in order to understand 
the different conclusions reached.

The issue of the dominant  area law behaviour 
arises for any gauge group $SU(N)$. Therefore  
it is also interesting  to study  $SU(4)$ and test whether the
SU(4) results corroborate the conclusions reached in SU(3).
As for $SU(3)$, we choose
lattice geometries which maximize the difference between the
two Ans\"atze which for $SU(4)$ is at the 20\% level.

As yet a further check, we evaluate the gauge-invariant nucleon wave function, using 
a density insertion for each quark line, and examine
whether it is better described according to the $Y$- or 
$\Delta$- law~\cite{AFT2}.

\section{Baryon Wilson loop}
For the static $q\bar{q}$ potential the appropriate operator is the standard
Wilson loop. For the static baryon potential 
in $SU(N)$, the corresponding operator is
 constructed by creating a gauge invariant 
$N$ quark state at time $t=0$ which is annihilated at a later time $T$.
Explicitly for  $SU(3)$,
the baryon Wilson loop, $W_{3q}$, shown in Fig.~1, is given by
\small
\be
\frac{1}{3!}\epsilon^{abc}\epsilon^{a'b'c'} U({\bf x,y},1)^{aa'}
        U({\bf x,y},2)^{bb'}U({\bf x,y},3)^{cc'}
\ee
 
\normalsize
\noindent
where 

\vspace*{-0.5cm}

\be
U(x,y,j)=P\exp\left[ig\int_{\Gamma(j)} dx^{\mu}A_\mu(x)\right] \quad,
\ee
$P$ denotes path ordering and $\Gamma(j)$ is the path from $x$ 
to $y$ for quark line $j$.

The $N$-quark potential is then extracted
from the long time behaviour of the Wilson loop:

\vspace*{-0.3cm}

\be
V_{Nq}=-\lim_{T \rightarrow \infty} \frac{1}{T} \ln \langle W_{Nq} \rangle \quad.
\ee


\begin{figure}[h]
\epsfxsize=5.5truecm
\epsfysize=5.5truecm
\mbox{\epsfbox{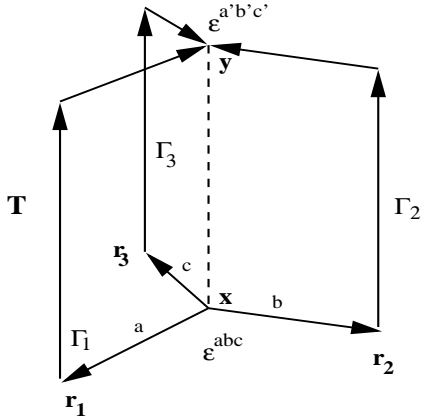}}
\vspace*{-1.5cm}
\caption{The baryonic Wilson loop in $SU(3)$. 
The quarks are located at positions ${\bf r}_1,{\bf r}_2$ and ${\bf r}_3$.}
\vspace*{-1.cm}
\label{Fig1}
\end{figure}

\section{Geometries in $SU(3)$ and $SU(4)$}
Two Ans\"atze exist in the literature regarding the area law behaviour
of the baryon Wilson loop:
\begin{itemize}
\item The $Y$-Ansatz: \\
In the strong coupling limit,
 minimization of 
the static energy amounts to giving the shortest length, $L_Y$,
 to the flux tubes joining the quarks.
For $SU(3)$, this is realized in general if
the three flux tubes meet at an interior point~\cite{CKP},
known as the Steiner point, where their mutual angles are $120^{0}$.
[If one of the angles of the triangle formed by the three quarks exceeds
$120^{0}$, the flux tube coming from that summit has length zero, and
the other two flux tubes meet there.]
Time evolution of this state in the general case produces 
a three-bladed area similar to Fig.\ref{Fig1}, known as the $Y$- area law.

For $SU(4)$ we have more possibilities.
Minimization of the static energy leads to the two stationary solutions
shown in Fig.~\ref{fig:BWL su4}, namely
 one configuration with a single Steiner point ($X$-Ansatz), and one with
two Steiner points A and B ($Y$-Ansatz).
The double string between the 
two Steiner points has tension 1.357(29) times greater\cite{Teper2}
than the other four, single strings.
If we neglect this difference for simplicity, then 
the $Y$-Ansatz always has lower energy than the $X$-Ansatz.
Here, we make no attempt to distinguish between the $Y$- and $X$-
Ans\"atze, and we assume that the double string has the same
tension as the single strings. Since this assumption has the 
effect of reducing the potential of the $Y$-Ansatz, which is itself lower
than in the $X$-Ansatz, it turns out to have no bearing on our
conclusions. 
In contrast to $SU(3)$ where
 for any given location of the three quarks, the Steiner
point and therefore the $Y$-Ansatz energy  can be computed analytically,
in $SU(4)$, the two Steiner points in the $Y$-Ansatz
are obtained by an iterative numerical procedure.

\begin{figure}
\vspace*{-0.3cm}
\epsfxsize=5.5truecm
\epsfysize=5.5truecm
\mbox{\epsfbox{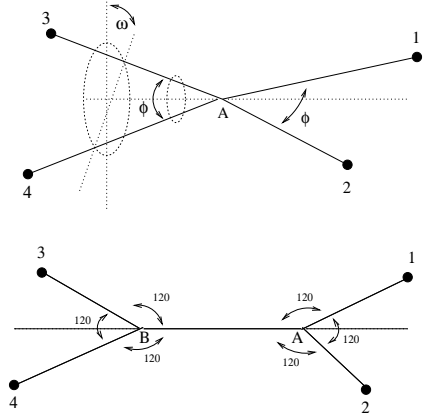}}
\vspace*{-0.5cm}
\caption{The flux tubes joining four quarks. The quarks are located
at positions ${\bf r}_1,{\bf  r}_2, {\bf r}_3$ and ${\bf r}_4$. 
The upper graph shows the local minimum of the energy
 with one Steiner point A, and
the lower is the minimum with two Steiner points A and B.}
\vspace*{-0.5cm}
\label{fig:BWL su4}
\end{figure}

\item The $\Delta$-Ansatz: \\
The second possibility for the relevant area 
dependence of the baryonic Wilson loop, proposed in ref.~\cite{Cornwall},
is that it is given by the sum
of the minimal areas $A_{ij}$ spanning quark lines $i$ and $j$. 
Because of its shape in $SU(3)$, this Ansatz is known as
the $\Delta$- area law. Therefore, we denote by $L_{\Delta}$ the total
 length  of all interquark distances.

\end{itemize}

For $SU(3)$ the maximal difference of 15\% between the two proposed area laws
is obtained when the 3
 quarks form an equilateral triangle. 

\begin{figure}[h]
\vspace*{0.5cm}
\epsfxsize=5.5truecm
\epsfysize=5.0truecm
\mbox{\epsfbox{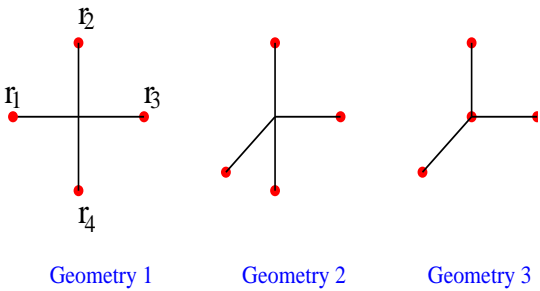}}
\vspace*{-2.5cm}
\caption{The geometries considered in  $SU(4)$. 
For geometry 1 the quarks are placed on a plane at equal distances from the
origin; for geometry 2 the positions of the four quarks are along
the three axes at (l,0,0), (0,l,0) (0,0,l), (0,0,-l);  and for geometry 3 at
(0,0,0), (l,0,0), (0,l,0), (0,0,l).}
\label{fig:geom_su4}
\vspace*{-0.8cm}
\end{figure}

For $SU(4)$ it
turns out that this relative difference
is maximal also for the configuration of maximal symmetry among the four quarks.
In this situation, where the quarks form a regular
tetrahedron, it reaches $21.96 \,\, \%$.
For programming convenience, we study instead the highly symmetric geometries
shown in Fig.~\ref{fig:geom_su4}, which all give a relative difference of
$\sim 20\%$.

\noindent

We compare our lattice results with the 
two expected forms of the baryonic potential which in  $SU(N)$  are
\begin{equation}
V_{Nq} = \frac{N}{2} \!V_0 
       -\frac{1}{N-1}\sum_{j<k}\frac{g^2 C_F}{4\pi r_{jk}} 
+ \sigma \left\{ \begin{array}{c} \!\!\frac{L_\Delta}{N-1} \\ L_Y \!\!\end{array}
                      \right\} 
\end{equation}
\noindent
with $C_F=(N^2-1)/2N$
and $\sigma$ the string tension of the $q\bar{q}$ potential. 
Note that in contrast to ref.~\cite{Takahashi} we do not allow $\sigma$ to vary.   
The factor of $1/(N-1)$ in the $\Delta-$ Ansatz makes
$L_\Delta/(N-1) < L_Y$ always.
In addition, we directly compare the three- or four-quark potential 
with the sum of
two-body potentials measured on the same gauge configurations, with no
adjustable parameters.

\section{Lattice techniques}
I.{\it Multi-hit procedure:} 
We carried out a comparison between the 
multi-hit procedure and 
hypercubic-blocking, as proposed in ref.~\cite{blocking}, for the time links.
In brief, hypercubic blocking smears the link using staples which all
belong to the $2^4$ hypercube surrounding it. Compared with the multi-hit
procedure, we find that hypercubic blocking tends to give
larger errors, especially for the large Wilson loops.
Since accuracy at large distances is our objective,
we have adopted the multi-hit procedure.
Moreover, for $SU(3)$ the group integral needed
 to obtain the  mean value of the 
link can be computed exactly \cite{Roiesnel}.
We note that the reduction factor in the statistical error from applying 
the multi-hit procedure to Wilson loops of time extent $t$ grows exponentially 
with $t$, and that the exponent is further multiplied by $3/2$ in the case of 
baryonic loops.

II.{\it Smearing Correlation matrix:}
We use a variational method to extract the groundstate potential.
For each quark configuration, we consider $M$ different levels of APE smearing, 
optimized as in ref.~\cite{alpha}, and construct an 
$M\times M$ correlation matrix $C(t)$ of Wilson loops of time extent $t$. 
We then solve the generalized eigenvalue problem
\be
C(t)v_k(t)=\lambda_k(t)C(t_0)v_k(t)
\ee
taking $t_0/a=1$.
We use two different methods to extract the groundstate energy:\\ 
1. In the first variant the potential is  extracted via
\be
aV_k={\rm lim}_{t\rightarrow \infty} -\ln
\left(\frac{\lambda_k(t+a)}{\lambda_k(t)}\right)
\ee
by fitting to the plateau. Note that with this method we also
estimate the energy of the first excited state.\\
2. In the second variant we consider the projected Wilson loops 
\be
W_P(t)=v_0^T(t_0)C(t)v_0(t_0)
\ee
and fit to the plateau value of  
$-{\rm ln}\biggl(W_P(t+1)/W_P(t)\biggr)$.
Both procedures gave consistent
results.
The energy of the
first excited state was used as a check in the extraction of 
the ground state, ensuring that the
contamination is less than $e^{-2}$ in the values considered for the plateau.

\section{Results}

For the baryonic loop in $SU(3)$ we
 used 220 configurations at $\beta=5.8$ and 200 at $\beta=6.0$ for a lattice 
of size $16^3\times 32$ from the NERSC archive.
For $SU(4)$ we generated 100 configurations at $\beta=10.9$
which gives a similar string tension $\sigma a^2$ as for $SU(3)$ at 
$\beta=6.0$.

\begin{figure}[h]
\vspace*{-2.cm}
\epsfxsize=5.3truecm
\epsfysize=6.0truecm
\mbox{\epsfbox{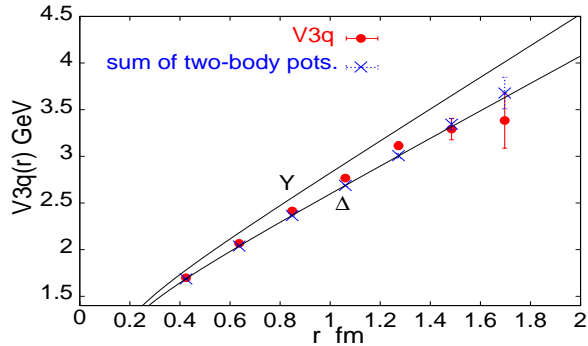}}
\vspace*{-0.5cm}
\caption{The static $SU(3)$ baryonic potential at $\beta=5.8$ (filled circles).
 The crosses
show the sum of the static $q\bar{q}$ potentials. The curves for the 
$\Delta$ and $Y$ Ans\"atze are also displayed. The quarks are located
at $(l,0,0)$, $(0,l,0)$, $(0,0,l)$ and $r = r_{12} = r_{13} = r_{23} = 
l \sqrt{2}\,\, $.}
\label{fig:beta58}
\vspace*{-0.5cm}
\end{figure}

\begin{figure}[h]
\vspace*{-2.5cm}
\epsfxsize=5.3truecm
\epsfysize=6.5truecm
\mbox{\epsfbox{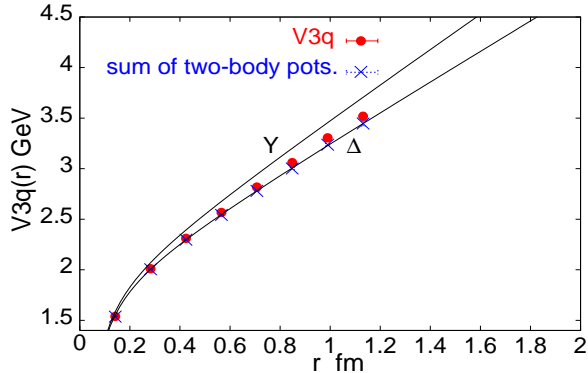}}
\vspace*{-0.5cm}
\caption{As Figure~\ref{fig:beta58} but for $\beta=6.0$.}
\label{fig:beta60}
\vspace*{-0.7cm}
\end{figure}

The results at $\beta=5.8$ and $6.0$ show reasonable scaling and are 
completely consistent with the sum of $q\bar{q}$ potentials extracted from
measurements on the same lattices, i.e. we find $V_{3q} \approx 3/2 \> V_{q\bar{q}}$, in
agreement with ref.~\cite{Bali}.

\begin{figure}[h]
\vspace*{-1.5cm}
\epsfxsize=5.3truecm
\epsfysize=6.5truecm
\mbox{\epsfbox{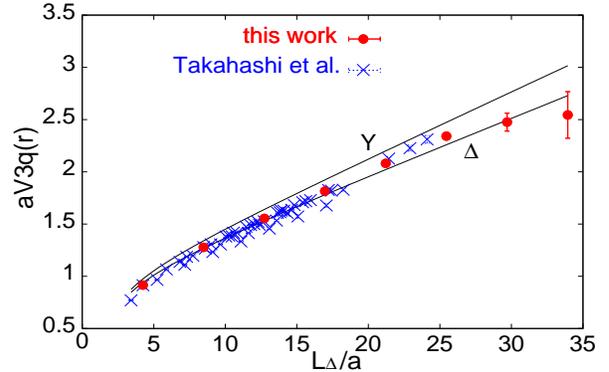}}
\vspace*{-0.5cm}
\caption{The static $SU(3)$ baryonic potential at $\beta=5.8$ for this
work (filled circles) and for ref.~\cite{Osaka2} (crosses) versus
the perimeter $L_\Delta$ of the triangle formed by the quarks.}
\label{fig:Japan}
\vspace*{-0.7cm}
\end{figure}

In Fig.~\ref{fig:Japan} we compare
our results with those obtained in ref.~\cite{Osaka2}. As it can  be seen
the two sets of data are in agreement for small loops 
which correspond to  the bulk of the
data of ref.~\cite{Osaka2}.
For larger loops, the few
 results of ref.~\cite{Osaka2} tend to lie above ours. 
The statistical errors on these data are not quoted but we expect them to be 
larger than ours, especially since the multi-hit procedure was not used. 
The analysis
of ref.~\cite{Osaka2} differs from ours in that we do not allow
the string tension to vary but take it from the fit to the $q\bar{q}$ 
potential. So in our approach having fixed
the $q\bar{q}$ potential there are no adjustable parameters that
enter in the two Ans\"atze.

\begin{figure}[h]
\epsfxsize=6.0truecm
\epsfysize=5.truecm
\centerline{\mbox{\epsfbox{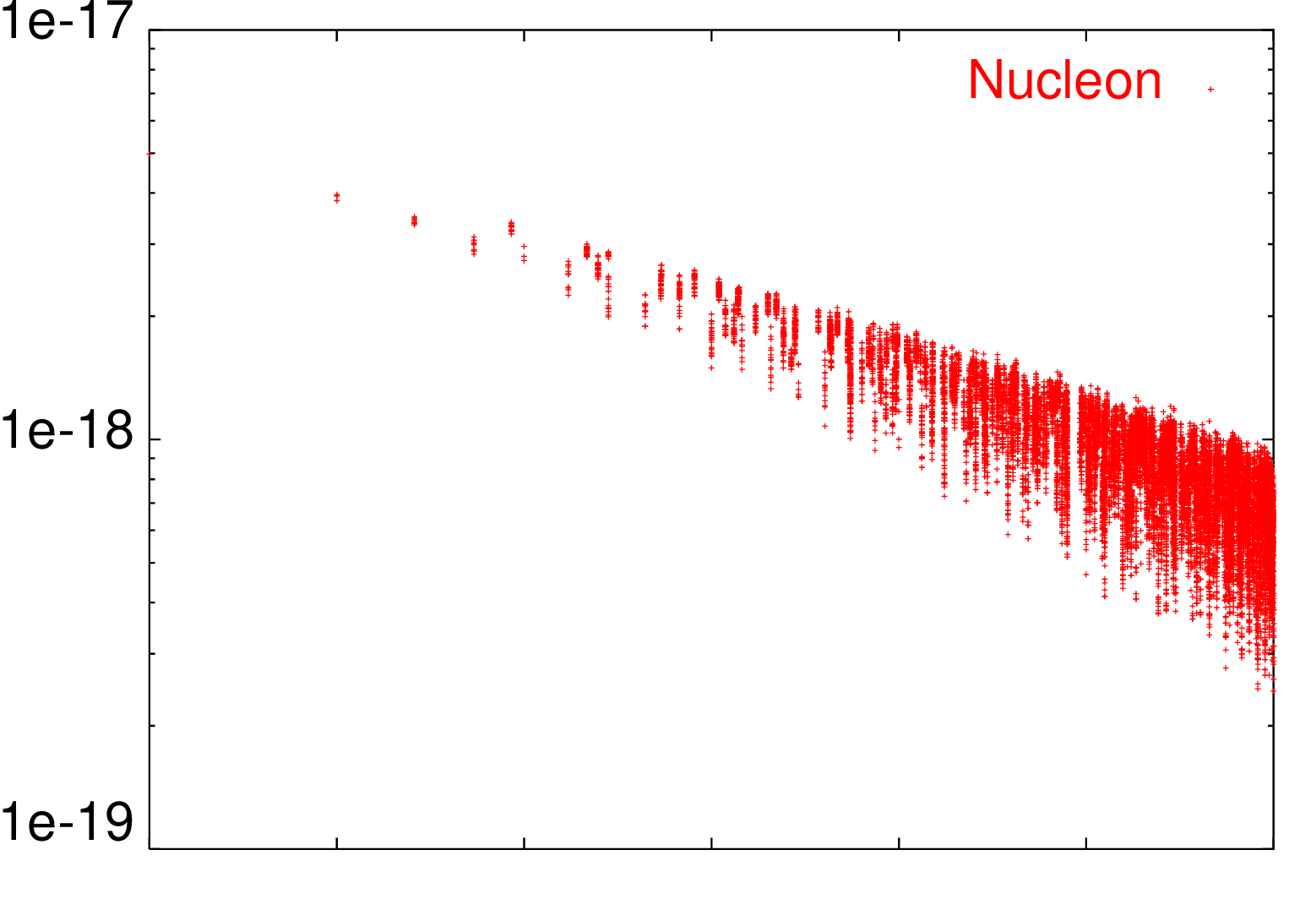}}}
\epsfxsize=6.truecm
\epsfysize=5.truecm
\vspace*{-0.5cm}
\centerline{\mbox{\epsfbox{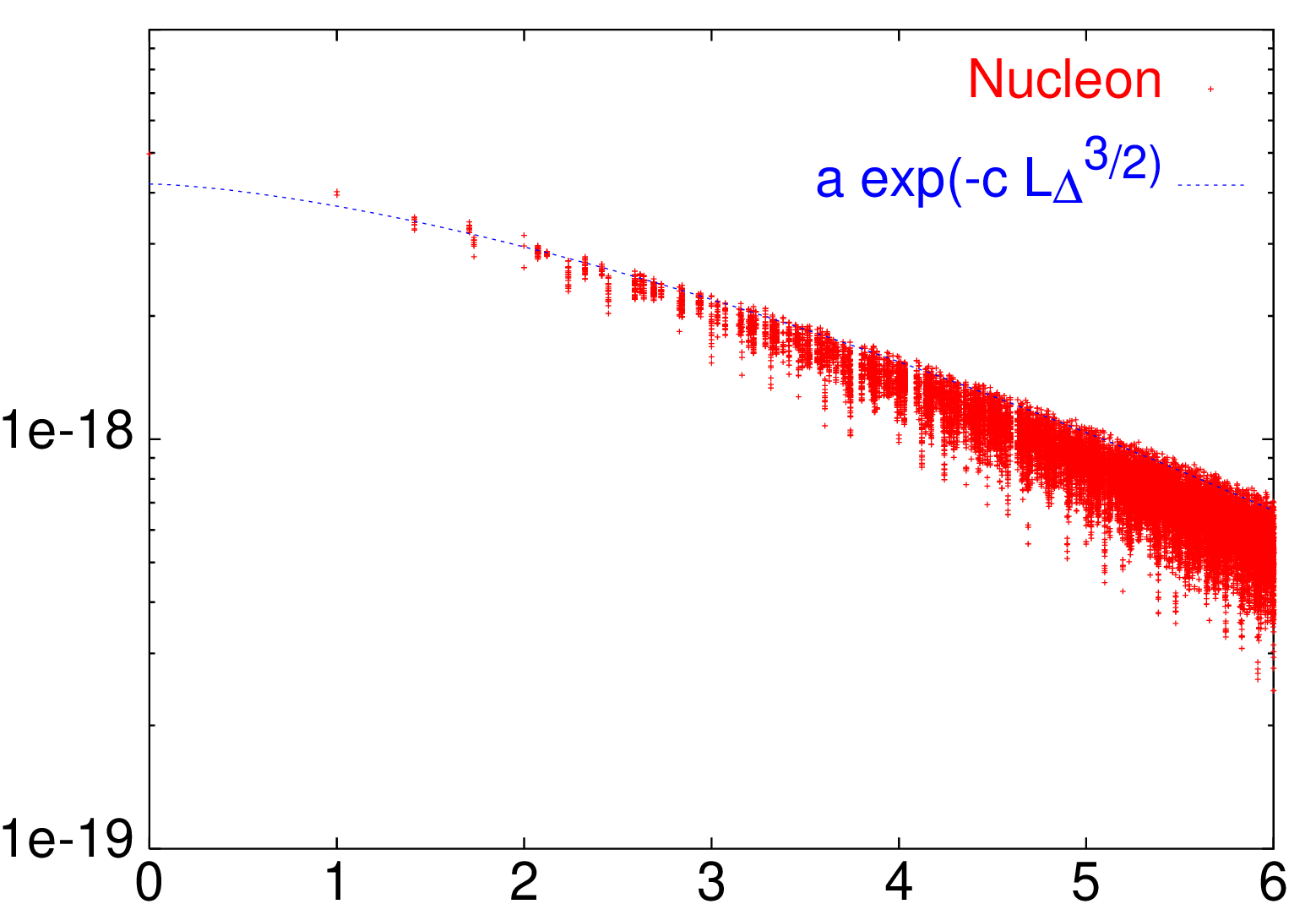}}}
\vspace*{-0.5cm}
\caption{Upper graph: The nucleon wave function versus $L_Y$.
Lower graph: The nucleon wave function versus $L_\Delta/2$, fitted by a linear potential ansatz. All quantities are in lattice units.}
\label{fig:wavefunction}
\vspace*{-0.7cm}
\end{figure}

\begin{figure}[h]
\vspace*{-1.5cm}
\epsfxsize=5.3truecm
\epsfysize=6.5truecm
\mbox{\epsfbox{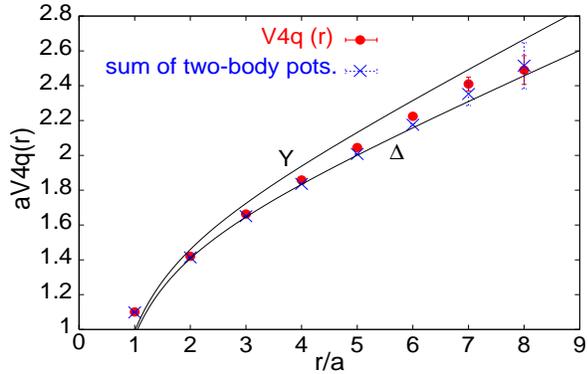}}
\vspace*{-0.5cm}
\caption{The static $SU(4)$ baryonic potential in lattice units for geometry 3.}
\label{fig:su4 geometry 3}
\vspace*{-0.5cm}
\end{figure}

Additional support for the $\Delta$-Ansatz is provided by
preliminary results on the gauge-invariant nucleon wave function, obtained by 
inserting three density operators at an intermediate time $t$ along the three quark lines
of a baryon propagator.
In Fig.~\ref{fig:wavefunction} we plot the wave function versus
$L_Y$ and $\L_\Delta/2$.
We observe a larger scatter when
 $L_Y$ is used as compared to  $L_\Delta$. 
The nucleon wave function  plotted versus $L_\Delta$ is displayed together
with a fit to the form $\exp(-c L_\Delta^{3/2})$, 
expected if the underlying potential is $\propto L_\Delta$.
As can be seen, this simple asymptotic 
form provides a remarkably good description of  
the nucleon wave function. 

Finally we display the results obtained in $SU(4)$ for geometry 3.
They give yet more support to the $\Delta$-area law for the baryon Wilson loop. The
other two geometries show the same behaviour.

\section{Conclusions}
Our results for the static three- and four-quark potential
in $SU(3)$ and $SU(4)$ are consistent with the sum of two-body potentials
and  inconsistent with the $Y-$ Ansatz
up to an interquark distance of about $0.8$~fm.
For larger distances, where our statistical and systematic errors both
become appreciable, there appears to be a small enhancement which could be
assigned to
the admixture of a many-body component. 
Nevertheless, for the distances
up to 1.2~fm that we were able to  probe in this work, 
the $\Delta$-area law gives the closest description of our data. 
More refined noise-reduction techniques
for the large loops will be needed
in order to clarify or rule out a many-body component 
at larger distances. Preliminary results on the nucleon wavefunction also 
support a potential in accord with the $\Delta$-Ansatz.

\end{document}